\documentclass[12pt, prd, showpacs]{revtex4}
\usepackage{amssymb}
\usepackage{amsmath}

\input{tcilatex}

\begin{document}

\title{Acceleration of particles by nonrotating charged black holes}
\author{Oleg B. Zaslavskii}
\affiliation{Kharkov V.N. Karazin National University, 4 Svoboda Square, Kharkov, 61077,
Ukraine}
\email{zaslav@ukr.net}

\begin{abstract}
Recently, in the series of works a new effect of acceleration of particles
by black holes was found. Under certain conditions, the energy in the centre
of mass frame can become infinitely large. The essential ingredient of such
effect is the rotation of \ a black hole. In the present Letter, we argue
that the similar effect exists for a nonrotating but charged black hole even
for the simplest case of radial motion of particles in the Reissner-Nordstr%
\"{o}m background. All main features of the effect under discussion due to
rotating black holes have their counterpart for the nonrotating charged ones.
\end{abstract}

\maketitle




\section{Introduction}

Recently, it was made an interesting observation that black holes can
accelerate particles up to unlimited energies $E_{cm}$ in the centre of mass
frame \cite{ban}. This stimulated further works in which details of this
process were investigated \cite{gp4} - \cite{ted} and, in particular was
found that the effect is present not only for extremal black holes but also
for nonextremal ones \cite{gp4}. These results have been obtained for the
Kerr metric (they were also extended to the extremal Kerr-Newman one \cite%
{kn} and the stringy black hole \cite{sen}). In the work \cite{oz1}
generalization of these observations was performed and it was demonstrated
that the effect in question exists in a generic black hole background (so a
black hole can be surrounded by matter) provided a black hole is rotating.
Thus, rotation seemed to be an essential part of the effect. It is also
necessary that one of colliding particles have the angular momentum $L_{1}=%
\frac{E_{1}}{\omega _{H}}$ \cite{oz1} where $E$ is the energy, $\omega _{H}$
is the angular velocity of \ a generic rotating black hole. If $\omega
_{H}\rightarrow 0$, $L_{1}\rightarrow \infty $, so for any particles with
finite $L$ the effect becomes impossible. Say, in the Schwarzschild
space-time, the ratio $E_{cm}/m$ ($m$ is the mass of particles) is finite
and cannot exceed $2\sqrt{5}$ for particles coming from infinity \cite{b}.

Meanwhile, sometimes the role played by the angular momentum and rotation,
is effectively modeled by the electric charge and potential in the
spherically-symmetric space-times. So, one may ask the question: can we
achieve the infinite acceleration without rotation, simply due to the
presence of the electric charge? Apart from interest on its own., the
positive answer would be also important in that spherically-symmetric
space-times are usually much simpler and admit much more detailed
investigation, mimicking relevant features of rotating space-times. As we
will see below, the answers is indeed "yes"! Moreover, in \cite{ban} - \cite%
{oz1} rotation manifested itself in both properties of the background metric
and in the nonzero value of angular momentums of colliding particles.
However, below we show that both manifestations of rotation can be absent
but nonetheless the effect under discussion reveals itself. This is
demonstrated for the radial motion of particles in the Reissner-Nordstr\"{o}%
m black hole, so not only $\omega _{H}=0$ but also $L_{1}=L_{2}=0$ for both
colliding particles. It is surprising that the effect reveals itself even in
so simple situation (which is discussed even in textbooks).

Formally, the results for the accleration of charged particle by the
Reissner-Nordstr\"{o}m black hole can be obtained from the corresponding
formulas for the Kerr-Newman metric. Although the Kerr-Newman metric was
discussed in \cite{kn}, only motion of uncharged particles with angular
momenta was analyzed, the metric being extremal, so there was no crucial
difference from the acceleration in the Kerr metric. However, now we are
dealing with the situation when a particular case is in a sense more
interesting than a general one since it reveals a qualitatively different
underlying reason of acceleration to infinite energies. We also discuss both
the extremal and nonextremal metrics.

\section{Basic formulas}

Consider the metric of the Reissner-Nordstr\"{o}m black hole%
\begin{equation}
ds^{2}=-dt^{2}f+\frac{dr^{2}}{f}+r^{2}d\omega ^{2}.
\end{equation}%
Here $d\omega ^{2}=\sin ^{2}\theta d\phi ^{2}+d\theta ^{2}$, $f=1-\frac{2M}{r%
}+\frac{Q^{2}}{r^{2}}$ where $M$ is the black hole mass, $Q$ is its charge.
The event horizon lies at $\ r=r_{H}=M+\sqrt{M^{2}-Q^{2}}$. \ Consider a
radial motion of the particle having the charge $q$ and rest mass $m$. Then,
its equations of motion read%
\begin{equation}
mu^{0}=m\dot{t}=\frac{1}{f}(E-\frac{qQ}{r})\text{,}  \label{u0}
\end{equation}

\begin{equation}
m^{2}\dot{r}^{2}=(E-\frac{qQ}{r})^{2}-m^{2}f\text{.}  \label{ur}
\end{equation}%
Here, $E$ is the conserved energy, dot denotes differentiation with respect
to the proper time $\tau $, $u^{\mu }$ is the four-velocity. In what
follows, we assume that the difference $E-\frac{qQ}{r_{H}}\geq 0$, so it is
positive for all $r>r_{H}$ (motion "forward in time") .

Let two particles (labeled by $i=1,2$) fall into the black hole, so $\dot{r}%
_{1}<0$, $\dot{r}_{2}<0$. The relevant quantity which we are interested in
is the energy in the centre of mass frame \cite{ban} - \cite{oz1} which is
equal to%
\begin{equation}
E_{cm}=m\sqrt{2}(1-u_{1\mu }u^{2\mu })\text{.}  \label{cm}
\end{equation}%
It follows from (\ref{u0}) - (\ref{cm}) that%
\begin{equation}
\frac{E_{cm}^{2}}{2m^{2}}=1+\frac{X_{1}X_{2}-Z_{1}Z_{2}}{fm^{2}}  \label{exz}
\end{equation}%
where 
\begin{equation}
X_{i}=E_{i}-\frac{q_{i}Q}{r}\text{, }Z_{i}=\sqrt{X_{i}^{2}-m^{2}f}\text{.}
\label{xz}
\end{equation}

\section{Limiting transitions for energy}

Now, we are going to examine what happens in different limiting transitions
which involve the near-horizon region where $f\rightarrow 0$.

1) Let $f\rightarrow 0$. Then, we obtain from the (\ref{exz}), (\ref{xz})
that 
\begin{equation}
\frac{E_{cm(H)}^{2}}{2m^{2}}=1+\frac{1}{2}[\frac{q_{2(H)}-q_{2}}{%
q_{1(H)}-q_{1}}+\frac{q_{1(H)}-q_{11}}{q_{2(H)}-q_{22}}]
\end{equation}%
where $q_{i(H)}\equiv \frac{E_{i}r_{H}}{Q}$. It is worth noting that, as $%
r_{H}\geq Q$ (for the definiteness, we take $Q>0$), the critical charge $%
q_{(H)}>E$. If \ a particle falls from infinity, $E>m\,$\ whence $q_{(H)}>m$%
, so a particle with the charge $q_{(H)}$ is overcharged in this sense.

2) If, say, $q_{1}=q_{1(H)}(1-\delta )$ with $\delta \ll 1$ and $q_{2}\neq
q_{2(H)}$, the energy $E_{cm(H)}\sim \frac{1}{\sqrt{\delta }}$ can be made
as large as one likes. Thus, we have that%
\begin{equation}
\lim_{q_{1}\rightarrow q_{1(H)}}\lim_{r\rightarrow r_{H}}E_{cm}=\infty .
\end{equation}

3) Let now $q_{1}=q_{1(H)}$ from the very beginning, $q_{2}\neq q_{2(H)}$.
Then, $X_{1}=E_{1}(1-\frac{r_{H}}{r})$. For the nonextremal horizon, in the
vicinity of the horizon the expression inside the square root is dominated
by the term \thinspace $-m^{2}f$ and becomes negative. This means that the
horizon is unreachable, so this case is irrelevant for our purposes.
Instead, let us consider the extremal horizon, $M=Q=r_{H}$, $f=(1-\frac{r_{H}%
}{r})^{2}$. After simple manipulations, we find that near the horizon, 
\begin{equation}
\frac{E_{cm}^{2}}{2m^{2}}=1+\frac{X_{2(H)}}{m^{2}(1-\frac{r_{H}}{r})}[E_{1}-%
\sqrt{(E_{1}^{2}-m^{2})}]+O((1-\frac{r_{H}}{r}))\text{.}
\end{equation}%
Thus, $E_{cm}$ diverges in the horizon limit:%
\begin{equation}
\lim_{r\rightarrow r_{H}}\lim_{q_{1}\rightarrow q_{1(H)}}E_{cm}=\infty \text{%
.}
\end{equation}

4) For completeness, we should also consider the case $q_{1}=q_{1(H)}$, $%
q_{2}=q_{2(H)}$ for the extremal horizon. Then, $X_{i}=E_{i}(1-\frac{r_{H}}{r%
})$, $Z_{i}=(1-\frac{r_{H}}{r})\sqrt{E_{i}^{2}-m^{2}}$ and we obtain that 
\begin{equation}
\frac{E_{cm}^{2}}{2m^{2}}=1+\frac{E_{1}E_{2}-\sqrt{E_{1}^{2}-m^{2}}\sqrt{%
E_{2}^{2}-m^{2}}}{m^{2}},
\end{equation}%
so the energy remains finite and this case is of no interest in the present
context.

\section{Limiting transitions for time and conditions of collision}

In the above consideration, we showed that the energy $E_{cm}$ can be made
as large as one likes provided $q_{1}\rightarrow q_{1(H)}$ and collision
occurs near the horizon. Meanwhile, it is also essential to be sure that
collision itself can be realized. Preliminarily, it can be understood that
this is indeed the case, by analogy with the Kerr case where this was issue
traced in detail \cite{ban} - \cite{ted}.

Consider what happens in more detail. Let at the moment of the coordinate
time $t=0$ particle 1 starts to move towards the horizon at the point $r_{i}$%
, at the later moment $t=t_{0}>0$ particle 2 does the same (the precedent
history of particles is unimportant). Then, the condition of collision at
the point $r=r_{f}$ reads%
\begin{equation}
t_{0}=t_{1}-t_{2}>0\text{, }t_{1}=\int_{r_{f}}^{r_{i}}\frac{drX_{1}}{f\sqrt{%
X_{1}^{2}-m^{2}f}}\text{, }t_{2}=\int_{r_{f}}^{r_{i}}\frac{drX_{2}}{f\sqrt{%
X_{2}^{2}-m^{2}f}}\text{.}  \label{t0}
\end{equation}%
To this end, it is sufficient (say, for $Q>0$) to take $q_{2}\leq q_{1}$, $%
E_{2}>E_{1}$. Then, $X_{2}>X_{1}$ for any $r$, $\ $and it is obvious that
indeed the time $t_{0}>0$.

Then $r_{f}\rightarrow r_{H}$, each of integrals in (\ref{t0}) diverges in
accordance with the well known fact that when the horizon is approached, the
time measured by clocks of a remote observer is infinite. Let us discuss
what happens to $t_{0}$ in the limiting situations 1)-4) discussed above

If we take the horizon limit 1) we find that $t_{0}$ is finite for $%
q_{1}\neq q_{1(H)}$, $q_{2}\neq q_{2(H)}$. Both proper times $\tau _{1}$, $%
\tau _{2}$ are also finite. If, afterwards, we consider $q_{1}=q_{1(H)}(1-%
\delta )$ with $\delta \ll 1,$ the time $t_{0}$ is still finite, the allowed
region for particle 1 near the horizon shrinks since the positivity of (\ref%
{ur}) entails that $r_{H}<r<r_{H}+A\delta ^{2}$ where $A=\frac{r_{H}^{2}}{2%
\sqrt{M^{2}-Q^{2}}}\left( \frac{E}{m}\right) ^{2}$. The proper time $\tau
_{1}\sim \delta ,\tau _{2}\sim \delta ^{2}$. In case 3) the horizon is
extemal and $q=q_{(H)}$. Then, one can obtain the exact explicit
expressions: 
\begin{equation}
t_{1}=\frac{E}{\sqrt{E_{1}^{2}-m^{2}}}[r+2r_{H}\ln (r-r_{H)}-\frac{r_{H}^{2}%
}{r-r_{H}}]_{r_{f}}^{r_{i}}
\end{equation}%
\begin{equation}
\tau _{1}=\frac{m}{\sqrt{E_{1}^{2}-m^{2}}}(r_{i}-r_{f}+\ln \frac{r_{i}-r_{H}%
}{r_{f}-r_{H}})\text{.}  \label{tau}
\end{equation}

If $r_{f}\rightarrow r_{H}$, $t_{1}\sim (r_{f}-r_{H})^{-1}\sim $ $t_{2}$.
The proper time $\tau _{1}$ diverges logarithmically, $\tau _{2}$ is finite,
so that the situation is very similar to the case of the extremal rotating
black holes (cf. \cite{gp4}, \cite{oz1}). Moreover, calculating the second
derivative $\ddot{r}$ from (\ref{ur}), one can see that in the case under
discussion both $\dot{r}$ and $\ddot{r}$ asymptotically vanish as the
particle approaches the horizon, so particle 1 halts in the sense that $r$
almost does not change (in terms of the proper distance $l$, the derivative$%
\frac{dl}{d\tau }$ is finite but $l$ itself diverges for the extremal
horizon). Correspondingly, particle 2 will inevitably will come up with a
slow falling particle 1 and will collide with it.

Thus, we checked that in all cases of interest particle 2 can indeed
overtakes particle 1, so collision will occur. This happens for a finite (or
even almost vanishing) interval of the proper time of particle 2 after the
start of motion in point $r_{i}$.

\section{Extraction of energy after collision}

Up to now, we discussed the effect of infinitey growing energy in the centre
of mass frame. Meanwhile, for observations in laboratory, it is important to
know what can be seen by an observer sitting at infinity. This poses a
question about the possibility of extraction of the energy after collision.
Below, we suggest preliminary analysis for the process near the horizon
similar to what has been carried out in Sec. 2 of \cite{gp4} for rotating
black holes.

Let two particles with energies $E_{1},E_{2}$ and charges $q_{1},q_{2}$
experience collision and turn into two other particles with energies $%
\varepsilon _{1},\varepsilon _{2}$ and charges $e_{1},e_{2}$. From the
conservation law we have for the energy, the radial momentum (\ref{ur}):and
the electric charge:%
\begin{equation}
E_{1}+E_{2}=\varepsilon _{1}+\varepsilon _{2}\text{,}
\end{equation}%
\begin{equation}
-[\sqrt{(E_{1}-\frac{q_{1}Q}{r})^{2}-m^{2}f}]+\sqrt{(E_{2}-\frac{q_{2}Q}{r}%
)^{2}-m^{2}f}]=[\sqrt{(\varepsilon _{1}-\frac{e_{1}Q}{r})^{2}-m^{2}f}]-\sqrt{%
(\varepsilon _{2}-\frac{e_{2}Q}{r})^{2}-m^{2}f}]
\end{equation}%
\begin{equation}
q_{1}+q_{2}=e_{1}+e_{2}.
\end{equation}%
$.$

The signs are chosen so, that before the collision both particles move
towards a black hole and after it particle 1 goes outside a black hole and
particle 2 goes inside. If collision occurs at the horizon $f=0$, the system
simplifies and one finds explicitly%
\begin{equation}
\varepsilon _{1}=\frac{Q}{r_{+}}e_{1}\text{, }\varepsilon
_{2}=E_{1}+E_{2}-\varepsilon _{1}\text{, }e_{2}=q_{1}+q_{2}-e_{1}\text{.}
\label{mu}
\end{equation}

From (\ref{mu}), we obtain the bound $\varepsilon _{1}\leq e_{1}$. If, say,
all particles have charges of the same order $q_{i}\sim e_{i}\sim q$ (i=1,2)
which remain the same after elastic scattering, then $\varepsilon
_{1}\lesssim q$. As far as the extraction of the energy is concerned, the
collision with the critical value of the charge $q_{H}$ of the falling
particle leading to the unbound energy in the centre of mass is not singled
out in this process. Thus, from the viewpoint of an observer at infinity,
one cannot gain much energy in elastic scattering. Rather, the main new
physical effects from the collision with unbound energy in the centre of
mass can be connected with new (yet unknown) physics at Planck scale due to
creation of new kinds of particles$.$

At the first glance, the bound for the extracted energy is similar to that
claimed in \cite{ted} for the rotating case, with $m$ replaced by $q$.
However, this bound was refuted in \cite{gp4} where no bound was found at
all. The difference between the situations considered here and in \cite{ted}%
, \cite{gp4} consists in that the angular momentum of a particle can be
taken arbitrarily large after collision while we assume that an electric
charge does not change (at least, significantly) after collisions.

\section{Critical electric charge and creation of pairs}

Up to now, our consideration was pure classical. Meanwhile, it is known that
in the electric field of a Reissner-Nordstr\"{o}m black hole, creation of
electron-positron pairs leads to diminishing of its electric charge \cite{E}%
. Does this process influence significantly the effect discussed in our
article? The pair production is energetically favorable, if $Q\geq \frac{\mu 
}{e}r_{H}c^{2}$ where $\mu $ is the electron mass, $e$ its electric charge
(in this section we restore explicitly the speed oflight $c$, the
gravitational constant $G$ and the Planck constant 
h{\hskip-.2em}\llap{\protect\rule[1.1ex]{.325em}{.1ex}}{\hskip.2em}%
). Meanwhile, in our consideration the charge of one of two colliding
particles should be close but slightly less that $q_{H}=\frac{mc^{2}}{Q}%
r_{H} $ (for simplicity, we assume that the energy $E\approx mc^{2}$ that
corresponds to the particle nonrelativistic at infinity). Therefore, if one
tries to apply our formulas to this case directly, it is seen that the
system is close to the threshold of pair production but is somewhat below
it. However, particle production comes into play in an indirect way.
Consider the collision of (quasi)classical heavy particles such that $\frac{%
\mu }{e}r_{H}c^{2}<Q<\frac{m}{q}r_{H}c^{2}$. Then, the creation of light
particles (electrons and positrons) will change the value of the black hole
charge and, thus, affect the critical value for heavy particles. As is known 
\cite{E}, the charge, independently on the exact initial value, falls off
rather rapidly to the value $Q_{1}=\frac{\pi \mu ^{2}c^{3}r_{H}^{2}}{e\text{%
h{\hskip-.2em}\llap{\protect\rule[1.1ex]{.325em}{.1ex}}{\hskip.2em}%
}}$. This entails the grow of the electric charge needed for the effect of
acceleration under discussion to occur. We assume that our colliding
particles represent or consist of stable atoms, so it is natural to assume
that $q_{H}<137e=\frac{\text{%
h{\hskip-.2em}\llap{\protect\rule[1.1ex]{.325em}{.1ex}}{\hskip.2em}%
c}}{e}$ to avoid quantum instability of particles themselves. Then, after
the substitution of $Q_{1}$ into the formula for $q_{H}$ we obtain the
inequality%
\begin{equation}
M>\frac{m}{G}\left( \frac{e}{\mu }\right) ^{2}\approx 10^{42}m\text{.}
\end{equation}

It is seen that this inequality does not contain the Planck constant. If we
take, say, $m\sim 100m_{p}$, \ where $m_{p}$ is the proton mass, we have
that $M>10^{20}g$ that is not restrictive from the astrophysical viewpoint.

\section{Conclusions}

Thus, we reproduced all main features, existing for acceleration of
particles in rotating black holes: for the nonextremal horizon the energy in
the centre of mass frame is finite but can be made as large as one like, for
the extremal case with the critical value of charge $q_{(H)}$ it diverges.
In the latter case the proper time is also infinite, so in a sense the
situation resembles the remote singularity unreachable in a finite proper
time.

In our consideration, we neglect gravitational and electromagnetic radiation
and backreaction on the metric. The case of nonrotating black holes is
especially useful in a given context in that it seems to facilitate the
evaluation of the role of such effects which is a separate subject for
further investigation.

In the present work, we restricted ourselves by the Reissner-Nordstr\"{o}m
metric but it is clear from the method of derivation that effect should
persist in more general situation of \ a black hole surrounded by matter
(so-called "dirty" black hole). In this sense, the effect is as universal as
its counterpart for rotating black holes \cite{oz1}, provided one of
colliding particle has a special value of the electric charge.

In spite of the infinite energy of collision in the centre of mass frame,
the energy of charged particles remains bound after elastic scattering.
Therefore, in this context one can expect new physical effects detectable in
observations (at least, in principle) not from high energy bursts but,
rather due to new channel of reactions at Planck scale, entailing new
scenarios in particle physics and astrophysics.

I thank Yuri Pavlov for stimulating discussion and useful comments.

\end{document}